\documentstyle[graphicx, amssymb, amsmath, psfig, epsfig]{mn}

\title[Cosmology using 3D weak lensing]{Cosmological constraints from COMBO-17 using\\ 3D weak lensing}
\author[T. D. Kitching et al.]
       {T. D. Kitching\thanks{tdk@roe.ac.uk}$^{1}$,
	A. F. Heavens$^{1}$, A. N. Taylor$^{1}$, M. L. Brown$^{1}$,\\\\ 
{\rm \LARGE K. Meisenheimer$^{2}$, C. Wolf$^{3}$, M. E. Gray$^4$,
  D. J. Bacon$^{1}$}
\\
$^{1}$SUPA\thanks{The Scottish Universities Physics Alliance}, Institute
for Astronomy, University of Edinburgh, Royal Observatory, Blackford
Hill, Edinburgh, EH9 3HJ, U.K.\\
$^{2}$Max-Planck-Institut f\"ur Astronomie, K\"onigsstuhl 17, 69117
Heidelberg, Germany\\
$^{3}$University of Oxford, Denys Wilkinson Building, Department of Physics, Wilkinson Building, Keble Road, Oxford OX1 3RH, U.K.\\
$^{4}$School of Physics and Astronomy, University of Nottingham, Nottingham,
NG7 2RD, U.K.}

\newcommand{\be}{\begin{equation}}
\newcommand{\ee}{\end{equation}}
\newcommand{\ba}{\begin{eqnarray}}
\newcommand{\ea}{\end{eqnarray}}
\newcommand{\nn}{\nonumber \\}

\newcommand{\lgl}{\langle}
\newcommand{\rgl}{\rangle}

\newcommand{\br}{\mbox{\boldmath $r$}}
\newcommand{\ux}{\mbox{\boldmath{$x$}}}
\newcommand{\uy}{\mbox{\boldmath{$y$}}}

\newcommand{\D}{\mbox{\boldmath $D$}}

\newcommand{\bell}{{\mbox{\boldmath{$\ell$}}}}
\newcommand{\edth}{\,\eth\,}

\def\gs{\mathrel{\raise1.16pt\hbox{$>$}\kern-7.0pt %
\lower3.06pt\hbox{{$\scriptstyle \sim$}}}}         %
\def\ls{\mathrel{\raise1.16pt\hbox{$<$}\kern-7.0pt %
\lower3.06pt\hbox{{$\scriptstyle \sim$}}}}         %

\begin{document}

\maketitle

\begin{abstract}
We present the first application of the 3D cosmic shear method
developed in Heavens et al. (2006) and the geometric shear-ratio
analysis developed in Taylor et al. (2006), to the COMBO-17 data
set. 3D cosmic shear has been used to analyse galaxies with 
redshift estimates from two random COMBO-17 fields covering $0.52$ square
degrees in total, providing a conditional
constraint in the ($\sigma_8$, $\Omega_m$) plane as
well as a conditional constraint on the equation of state of dark
energy, parameterised by a constant $w\equiv p_{\rm de}/\rho_{\rm
  de}c^2$. The 
($\sigma_8$, $\Omega_m$) plane analysis constrained the relation
between $\sigma_8$ 
and $\Omega_m$ to be $\sigma_8(\Omega_m/0.3)^{0.57\pm
  0.19}=1.06^{+0.17}_{-0.16}$, in agreement with a 2D
cosmic shear analysis of COMBO-17. The
3D cosmic shear conditional constraint on $w$ using the two random fields is
$w=-1.27^{+0.64}_{-0.70}$. The 
geometric shear-ratio analysis has been applied to the A901/2 field,
which contains three 
small galaxy clusters. Combining the analysis from the
A901/2 field, using the geometric shear-ratio analysis, and the two
random fields,  
using 3D cosmic shear, $w$ is conditionally constrained to
$w=-1.08^{+0.63}_{-0.58}$. The errors presented in this paper are shown to
agree with Fisher matrix predictions made in Heavens et al. (2006) and
Taylor et al. (2006). When these methods are applied to large
datasets, as expected soon from surveys such as Pan-STARRS and
VST-KIDS, the dark energy equation of state could be constrained to an
unprecedented degree of accuracy.
\end{abstract}


\section{Introduction}
This paper presents the first application of 3D weak lensing
techniques developed in Heavens (2003), Heavens et al. (2006), Jain \&
Taylor (2003) and Taylor et al. (2006) to data. The data used is the
COMBO-17 survey (Wolf et al. 2001; Wolf et al., 2004) which is a
multi-band photometric survey with 
exceptional image quality and is ideal for a 3D weak lensing study. The
power of these methods is cosmological parameter estimation, focussing
especially on measuring the equation of state of dark energy which
appears to be responsible for the acceleration of the Universe.

The case that the isotropic expansion of the
Universe is accelerating is now convincing. The acceleration can be
attributed to the 
effect of a negative pressure component, dark energy, which accounts
for approximately $70\%$ of the mass-energy of the Universe. However the
identity of dark energy is entirely unknown. The nature of dark energy
may be ultimately determined by establishing its equation of state,
parameterised by  
\be
w\equiv p_{\rm de}/\rho_{\rm de}c^2.
\ee
A cosmological constant
has $w=-1$, a dynamical dark energy such as quintessence may
have $w\not=-1$. Note that in general $w$ may be a function of
redshift, $z$. This study attempts to use $3$D weak lensing to
constrain $w$ (where we have assumed that $w$ is a constant), for the
first 
time. It is based on  
only $0.78$ square degrees of data, and is essentially a proof of
concept in preparation for much larger surveys such as VST-KIDS,
Pan-STARRS (Kaiser, 2005) or the Dark Energy Survey (Wester, 2005)
which could lead to very accurate measurements of $w$ and its redshift
evolution.  

We also present constraints on the amount of matter in the Universe
$\Omega_m$ and the clustering of matter, parameterised by $\sigma_8$, the
rms of the fractional mass density fluctuations in spheres of radius
$8h^{-1}$ Mpc. Weak
lensing has already proven to be a powerful probe of both $\Omega_m$
and $\sigma_8$, using 2D weak lensing techniques. We show that a fully
3D shear analysis can also 
place tight constraints on the matter content and clustering. 

Other surveys have used weak lensing
data to constrain cosmological parameters using 2D and tomographic
tests. Most recently Semboloni et
al. (2006), Hoekstra et al. (2006), Schrabback et al. (2006) and
Hetterscheidt et al. (2006) 
have all constrained $\sigma_8$, $\Omega_m$ and
$\Omega_{de}$, though all these surveys cover a much larger area than
COMBO-17 . Hoekstra et al. (2006) use the first data release of the
CFHTLS Wide survey, which 
covers $22$ square degrees in the $i'$-band, to constrain
$\sigma_8=0.85\pm 0.06$ for $\Omega_m=0.3$, the full CFHTLS survey
will cover $170$ square 
degrees in $5$ photometric bands. Semboloni et al. (2006)
use the CFHTLS Deep data which 
covers $2.34$ square degrees in ($u^*$, $g'$, $r'$, $i'$, $z'$) to 
constrain $\sigma_8=0.89\pm 0.06$ 
for $\Omega_m=0.3$. Semboloni et al. (2006) also combine with Hoekstra
et al. (2006) to place an upper bound on $w$, 
marginalizing over $\Omega_m$, of $w<-0.8$. 

COMBO-17 has the best and most
reliable photometric redshifts to date, due to the large number of
bands, and so is ideal as a survey to test the 3D weak lensing
constraints. COMBO-17 is then ideal for this proof of concept, however
as shown in Heavens et al. (2006) and Taylor et al. (2006) when much
larger survey areas are available, a $5$-band large area survey could
constrain the dark energy equation of state much better than
correspondingly smaller area $17$-band survey. 

The first 3D weak lensing technique used in this paper is a 3D
cosmic shear analysis which 
analyses the full 3D 
cosmic shear field using a spherical harmonic expansion, proposed in
Heavens (2003) and developed in
Castro et al. (2005) and Heavens et al. (2006).
The second technique, suggested by Jain \& Taylor (2003), is a
geometric shear-ratio analysis 
which takes the ratio of tangential shears around galaxy
clusters, developed in Taylor et al. (2006). Both of these 
analyses are applied to the data available.  We
use 3D cosmic shear to place conditional constraints on $w$ and to
place conditional constraints in the ($\sigma_8$, $\Omega_m$) plane
using the two random fields. The A901/2 field is centered on a known
supercluster, so we use the geometric test on this field to place a conditional
constraint on the dark energy equation of state parameter $w$. This
separation of the data allows for a combination of the two methods in
the constraint of $w$. 

The results of this paper are a proof of method for these
techniques. Heavens et al. (2006) and Taylor et al. (2006) show that
in order to constrain the dark energy equation of state to $\Delta
w\approx 0.01$ large and deep photometric surveys will be needed. The
errors on the results in this paper are compared to predictions made
using the Fisher 
matrix formalism used in Heavens et
al. (2006) and Taylor et al. (2006). Brown et al. (2003) have already
applied 2D 
weak lensing to the 
COMBO-17 data set to constrain the ($\sigma_8$, $\Omega_m$) plane.
The results presented in this paper using the fully 3D cosmic shear analysis
on the same data set, should be in approximate agreement (but do
slightly better than) a 2D analysis. It should also be noted that in
order to constrain the dark 
energy equation of state one necessarily needs 3D methods. To
determine whether dark energy is a field or a manifestation of of
modified gravity (e.g. Ishak et al., 2006), one needs methods can
probe the expansion history of Universe, as is the case with the 
geometric shear-ratio analysis, or both the expansion history and the
growth of structure as is the case in the 3D cosmic shear analysis.

The structure of this paper is as follows: firstly the COMBO-17 data
set will be introduced and discussed in Section \ref{COMBO-17 Survey},
the application of the 3D cosmic shear analysis to the two random fields  
will be presented in Section \ref{The 3D Spectral Test},
the geometric shear-ratio analysis and the results of applying the
method to the 
A901/2 field will be presented in Section \ref{The Geometric Ratio
  Test}. The
constraints from the A901/2 field (using the geometric shear-ratio
analysis) and 
the constraints from the two random fields (using the 3D cosmic shear
analysis) will be
combined in Section \ref{A Combined Constraint}. Conclusions will be
presented in Section \ref{Conclusion}. 

\section{The COMBO-17 Survey}
\label{COMBO-17 Survey}
The COMBO-17 survey is a 17-band photometric redshift survey with
gravitational lensing quality R-band data (Wolf et al., 2001; Wolf et
al., 2004). The
survey consists of five fields each covering $0.26$ square degrees. All
of the fields were  
observed using the Wide-Field Imager (WFI) at the MPG/ESO 2.2m
telescope on La Silla in Chile, with a $4\times2$ array of
$2048\times 4096$ pixel CCDs, each pixel subtending 0.238
arcseconds. 

In this paper we will use three of the COMBO-17 fields, which were
observed and reduced earlier than the remaining two, and for which
there are redshift estimates and a shear catalogues available.
One of the fields used, the A901/2 field, is centred on the Abell 901/2
supercluster which has previously been analysed in 2D by Gray et
al (2002) and in 3D by Taylor et al. (2004). The A901/2 supercluster  
consists of three smaller clusters; A901a, A901b, A902, all at a
redshift of $z\approx 0.16$. It should be noted that supercluster
refers to a `web of clusters', the individual clusters are much smaller
$\sim 10^{14} M_{\odot}$ (see Taylor et al., 2004) than large strong
lensing clusters for example A1689. For an individual cluster the
fractional error on $w$ should decrease 
as the mass of the cluster increases. However in a large area survey there
should be many more low and medium mass clusters than large clusters
so that the constraint on $w$ is dominated by the numerous medium mass
clusters; for a detailed discussion see Taylor et al. (2006). 

The COMBO-17 survey also observed a randomly selected area of sky, and a
relatively empty, but well observed area. 
The S11 field was a 
randomly selected area of sky, that contains a moderately large
cluster Abell 1364 at a redshift of $z\approx 0.11$. The CDFS field was
chosen to overlap the Chandra Deep Field South, a relatively
`empty' region of sky containing no significant galaxy
clusters. We only used galaxies with reliable photometric redshifts and
with an $R$ magnitude of $R\leq 24$.

\subsection{Photometric Redshifts}
\label{Photometric Redshifts}
Each of the COMBO-17 fields was observed in 17 different filters,
with the intention of obtaining object classification and accurate
photometric redshifts. In order to provide reliable redshifts, the
filter set included five broad-band filters ($UBVRI$) and 12
medium-band filters from 350 to 930 nm. This observing strategy
allows simultaneous estimates of Spectral Energy Distribution
(SED) classifications and photometric redshifts from
empirically-based templates.  Wolf et al. (2001) describe in detail
the photometric redshift estimation methods used to obtain typical
accuracies of $\sigma_z\approx 0.05$ for galaxies throughout $0<z<1$.
We fit an empirical line to the data such that
\be
\label{photozeq}
\sigma_z(z)=0.03(1+z)^{1.5},
\ee
this $\sigma_z(z)$ is used in the likelihood analysis. It should be
noted that the parameter constraints are not sensitive to the exact
functional form of $\sigma_z(z)$. 
%\begin{figure}
%\psfig{figure=errors.eps,width=\columnwidth,angle=0,clip=}
%\caption{The photometric redshift errors as a function of redshift for
%  the COMBO-17 survey. The green (light grey) line shows the fit used
%  in this analysis $\sigma_z(z)=0.03(1+z)^{1.5}$.}\label{errors}
%\end{figure}

\subsection{Shear Measurements}
Throughout the observing campaign the $R$ filter was used in best
seeing conditions, in order to provide a deep $R$-band image from
which to measure the gravitational shear. Gray et al. (2002)
discuss the procedure used to reduce the $R$ band imaging data,
which totalled 21 hours for the three fields used. As described by
Gray et al. (2002) and Brown et al. (2003) the 352 individual 
chip exposures for each field were registered using linear
astrometric fits, with a 3$\sigma$
rejection of bad pixels and columns.

The Kaiser, Squires \& Broadhurst (1995; KSB) weak lensing measurement
method was applied, using the {\tt imcat} shear analysis package, to
the reduced images (see Gray et al., 2002; Brown et al., 2003). 
This resulted in a catalogue of galaxies with centroids and shear
estimates throughout the fields, corrected for the effects of
anisotropic smearing and point spread function (PSF) circularisation. 
The photometric redshift estimates were appended to this catalogue for
each galaxy from the standard COMBO-17 analysis of the full
multi-colour dataset. Of the 37,243 galaxies in the shear
catalogue, 36\% have a reliable photometric redshift, the remainder
being fainter than the $R=24$ reliability limit of the redshift
survey. The requirement for the 3D lensing study, that the
redshift of each galaxy be known, clearly results in an immediate
reduction of available galaxies. It is apparent that most of
the background sample is composed of galaxies that are small, and
fainter than the magnitude limit of the redshift survey. These
catalogues are the raw data used in this analysis. Brown et al. (2003)
also include galaxies without assigned redshifts into their analysis,
this can also be done in the case of 3D weak lensing however since
this paper is a proof of concept for the 3D weak lensing methods the
galaxies without redshifts will be left out of this analysis.

\section{The 3D Cosmic Shear Analysis}
\label{The 3D Spectral Test}
We have applied the 3D cosmic shear method to the CDFS and S11 fields of
COMBO-17 in order to constrain $w$ and jointly constrain
$\sigma_8$ and $\Omega_m$. Bacon et
al. (2005) analysed COMBO-17 using a real-space 3D cosmic shear method
to constrain the evolution of dark matter clustering. The results
presented in this Section are  
based on the methods outlined in Heavens et al. (2006).

\subsection{3D Cosmic Shear Likelihood}
The statistics we choose to work with are the transforms of the galaxy
ellipticities. The transforms are are defined, for a given
radial $k$-mode and angular $\bell$-mode, by summing over all galaxies
$g$, each at a redshift $z$ and angular position $\btheta_g$, 
in a given field catalogue 
\be
\label{estimators}
\hat\gamma_i(k,\bell)=\sqrt{\frac{2}{\pi}}\sum_{g}
e^g_i k j_{\ell}(kr^0_g)X_{\bell} \, {\rm e}^{-i\bell.\btheta_g}W(z).
\ee
The hat indicates that these are estimators of the transform of the
shear field, but the relationship is not direct as the number density
of sources is non-uniform. 

$g$ refers to the galaxies in the sample, $e_i^g$ is the $i$ component
of the complex ellipticity of the galaxy related to the weak
shear $\gamma_i^g$ and the intrinsic ellipticity of the galaxy by
$e_i^g=e_i^g({\rm intrinsic})+\gamma_i^g$. $r^0_g$ denotes the comoving 
distance to a galaxy calculated from the photometric redshift of the
galaxy by assuming a fiducial cosmology. The $j_{\ell}(z)$ are spherical Bessel
functions. $W(z)$ is a weighting
function which we set to $W(z)=1$ for the remainder of this
paper; for an investigation of the effect of changing the weighting
scheme see Heavens et al. (2006). The $X_{\bell}$ factor is introduced as a result 
of relating the potential to the shear field 
$\gamma(\br)=\frac{1}{2}\edth\edth\phi(\br)$ (see Heavens et al., 2006) 
and is given by 
\begin{equation}
\label{SPEeq7}
X_{\bell} \equiv \frac{(\bell_y^2-\bell_x^2)+2i\bell_x\bell_y}{\bell^2}.
\end{equation}

The expansion using spherical Bessel functions is natural 
for a flat universe. For non-flat universes, the appropriate functions
are ultra-spherical Bessel 
functions $\Phi^{\ell}_{\beta}(y)$, but in the $\ell\gg 1$ and 
$k\gg$ (curvature scale)$^{-1}$ r\'egime these are well approximated
by ordinary Bessel functions $\Phi^{\ell}_{\beta}(y)\rightarrow
j_{\ell}(kr)$ (Abbott and Schaefer, 1986; Zaladarriaga and Seljak,
2000). 

For each $e^g_i$ component there is a real and imaginary estimator
(via the $X_{\bell} \, {\rm e}^{-i\bell.\btheta_g}$ factor)  so
that the 
whole data vector used in the likelihood analysis consists of four
independent vectors at each $k$ and $\bell$:
$\hat\gamma_1^{R}$, $\hat\gamma_1^{I}$, $\hat\gamma_2^{R}$,
$\hat\gamma_2^{I}$, 
where the $R$ superscript denotes the real part of the $\hat\gamma_i$
estimator and 
$I$ the imaginary part. Note that this is for a given $\bell$-mode. 

The fiducial cosmology, denoted by the superscript $0$,
is in this case chosen to have 
$\Omega_m=0.3$, $\Omega_{de}=0.7$,  $\Omega_{b}=0.04$, $h\equiv
H_0/100$ kms$^{-1}$Mpc$^{-1}=0.71$,
$\sigma_8=0.8$, $w=-1.0$; we also set the scalar
spectral index to be $n_s=1.0$ and its running $\alpha_n=0.0$. 
The choice of the fiducial cosmology does not affect the results presented
in this paper, as long as the measured shear estimates and theoretical
calculations use the same fiducial cosmology to calculate the transform
coefficients. This fiducial cosmology simply acts, via the spherical
Bessel functions, to weight the shear values in a particular way. The
cosmological dependence comes from the shear values themselves,
$\gamma_i^g$, 
the cosmological dependence of the calculated covariance matrices come
from modelling the shear-shear covariance. We tested a variety of
fiducial models and the results were indeed unaffected.

Note that the average value of $\hat\gamma_i(k,\bell)$ is zero, so
that information on the cosmological 
parameters comes from the dependence of the signal part of the
covariance matrix $C$ i.e. we adjust the parameters until the {\em
covariance} of the model matches that of the data.  This was the
approach of Heavens and Taylor (1995); Ballinger, Heavens and Taylor
(1995); Tadros et al. (1995); Percival et al. (2004) in analysis of
large-scale galaxy data. The details of the covariance matrix
derivation are given in Heavens et al. (2006), where the covariance
matrix is given as the sum of signal and noise terms $C=S+N$. The
signal part of the covariance of $\gamma(k,\ell)$ 
for a survey of size 
$\Delta\theta\times\Delta\theta$ can be written as
\ba
\label{SMA1}
S&=&\langle\gamma(k,\bell)\gamma^*(k',\bell')\rangle_S\nn &=&
\int\frac{d^2\tilde{\bell}}{(2\pi)^2}
Q(\bell,\bell',\tilde{\bell},k,k') |X_{\tilde{\bell}}|^2\nn
&&\int_{-\Delta\theta/2}^{\Delta\theta/2}d^2\btheta{\rm
  e}^{-i(\bell-\tilde{\bell}).\btheta}
  \int_{-\Delta\theta/2}^{\Delta\theta/2}d^2\btheta'{\rm
  e}^{-i(\bell'-\tilde{\bell}).\btheta'}.
\ea
The $Q$ matrix is given by 
\be
\label{SMA2}
Q(\bell,\bell',\tilde{\bell},k,k')={9 \Omega_m^2 H_0^4 \over
  4\pi^2 c^4}\int {d\tilde k\over \tilde k^2}\,
G(\bell,\tilde{\bell},k,\tilde k)
G(\bell',\tilde{\bell},k',\tilde k)
\ee
where
\be
\label{SMA3}
G(\bell,\tilde{\bell},k,\tilde k)\equiv k\int dz\,dz_p\,\bar n_z(z_p)p(z_p|z)U(\tilde{\bell},r,\tilde k) j_\ell(kr^0)
\ee
where $z_p$ is an integral over redshift given an assumed
cosmology, and the integral over $z$ uses the cosmology to be
tested. $\bar n_z(z)$ is the predicted number density of objects as a
function of redshift which is measured from the survey in question.
$p(z_p|z)$ is a probability distribution in redshift which,
by convolving with the redshift distribution, takes into account the
uncertainty in redshift. We assume a Gaussian probability distribution
with the width being the measured photometric redshift as a function
of redshift (see Section \ref{Photometric Redshifts}).

The $U$ matrix used in equation (\ref{SMA3}) is  
\be
\label{SMA4}
U(\tilde{\bell},r,\tilde k)\equiv \int_0^r d\tilde r \,{F_K(r,\tilde r)\over
a(\tilde r)} \sqrt{P_\delta(k; \tilde r)} \, j_{\tilde{\ell}}(k \tilde
r),
\ee
where $P_\delta(k;r)$ is the matter power spectrum for the cosmology
to be tested. We compute the nonlinear power spectrum using the
fitting formulae 
of Smith et al. (2003), based on linear growth rates given by
Linder and Jenkins (2003). $F_K(r,r')=(1/r'-1/r)$ for a flat Universe (assumed
in this paper) and $a(r)$ is the dimensionless scale factor. 

The integrals over $\btheta$ in equation (\ref{SMA1}) can be evaluated so
that
\be
\label{SMA5}
S=\int\frac{d^2\tilde{\bell}}{(2\pi)^2}
Q(\bell,\bell',\tilde{\bell},k,k')
|X_{\tilde{\bell}}|^2{\cal F}(\bell,\bell',\tilde{\bell}),
\ee
where 
\ba
\label{SMA6}
{\cal F}(\bell,\bell',\tilde{\bell})\equiv\prod_{i=x,y}
\frac{4}{(\tilde{\bell}-\bell)_i(\tilde{\bell}-\bell')_i}
\sin\left[(\tilde{\bell}-\bell)_i\frac{\Delta\theta}{2}\right]\nn
\, \, \, \, \, \, \, \, \, \, \sin\left[(\tilde{\bell}-\bell')_i\frac{\Delta\theta}{2}\right]
\ea
$i=x,y$ represents the $x$ and $y$ components of a Cartesian
coordinate system for a survey. 

The shot noise part of the covariance matrix is calculated by assuming
a Poisson sampling an underlying smooth density field and is given by
\ba
\label{SPEeq27b}
N& = &\langle\hat\gamma_\alpha(k,\bell)\hat\gamma_\beta^*(k',\bell')\rangle_{SN}\nn
& = & \frac{\sigma_{\epsilon}^2\Delta\Omega}{4\pi^2} \int dz\, \bar n_z(z) k
j_\ell(kr^0)k'j_\ell(k'r^0)\,
\delta^K_{\alpha\beta}\delta^K_{\ell\ell'}
\ea
where
$\Delta\Omega=\Delta\theta\times\Delta\theta$. $\sigma_{\epsilon}$ is
measured from the data, for the CDFS field 
$\sigma_{\epsilon}=0.19$ and for the S11 field
$\sigma_{\epsilon}=0.22$.

For large surveys the calculation simplifies as ${\cal
  F}\rightarrow(\Delta\theta)^2\delta^K_{\ell\tilde{\ell}}
\delta^K_{\ell'\tilde{\ell}}$ (see equation \ref{SMA6}) and the
covariance matrix becomes diagonal in ($\bell$, $\bell'$). For
COMBO-17, the survey is small so 
that it is necessary to compute the integral in equation (\ref{SMA5})
accurately using the full ${\cal F}(\bell,\bell',\tilde{\bell})$. We
do this for the diagonal components of $Q$, but we make an
approximation by ignoring the off-diagonal components. Improvement
on this would involve computing the full covariance resulting 
in vast computational expense, which is not really warranted by the
size of the dataset. All correlations between $k$-modes, for
any given $\bell$-mode are fully taken into account.

We assume then that the distribution of $\bell$-modes can
be represented by a multivariate 
Gaussian. The likelihood function for a given $\bell$ and set of 
cosmological parameters $\{\theta_{\alpha}\}$ is given by 
\ba
-2 \ln L_{\ell}(\theta_{\alpha}|\D) &=&
\sum_{A=\{R,I\}}\sum_{i=\{1,2\}} N^A_i\ln(2\pi)\nn 
& &+\ln(|C_{i,\ell}(k,k')^{AA}|) \\
& &+\sum_{kk'} \hat\gamma^A_{\ell}(k)
(C^{-1}_{i,\ell}(k,k'))^{AA}\hat\gamma^{A,T}_{\ell}(k')\nonumber
\ea
where $A=\{R,I\}$ is a sum over the real and imaginary estimators and
$i=\{1,2\}$ is a sum over the $\gamma_1$ and $\gamma_2$ shear
components. $N^A_i$ is the number of $k$-modes in the $N^A_i\times
N^A_i$ covariance matrix $C_{i,\ell}(k,k')^{AA}$. The log-likelihood is
then summed over each independent 
$\bell=(\ell_x, \ell_y)$ mode. Note that since we have four
independent data vectors, two real and imaginary pairs, we only
investigate the range $\ell_x\geq 0$ to avoid double counting. 

The $\ell$ and $k$ ranges and resolutions used are as follows. In
integrating over $\tilde{\bell}$ in equation (\ref{SMA1}) we found that
the signal converges 
at $\Delta\tilde{\ell}_i=100$ and for the range
$(\ell_i-1500)<\tilde{\ell}_i<(\ell_i+1500)$ where $i=x,y$. The $k$
resolution used was $\Delta k=2\times 10^{-3}$ Mpc$^{-1}$, as it was
found that 
the signal part of the covariance matrix converges at $\Delta
k\approx (2\pi/r_{\rm max})\approx 2\times 10^{-3}$  Mpc$^{-1}$ where
$r_{\rm max}$ is the distance corresponding to a maximum redshift of
$z\approx 1$. The $k$ range used was $0.01< k < 1.5$ Mpc$^{-1}$. The
$\ell$ values available are constrained by the 
survey geometry, $\ell_i=\frac{2\pi n}{\Delta\Theta}$ where $n$ is an
integer; $\ell_1=671$. We tested the lower $\ell$ limit and found no
change in the cosmological constraints by using $\ell_1=700$
instead, showing that these results are robust to the details of the
lower $\ell$ range used. We use all  
modes with $|\bell|\leq 2500$ to avoid the highly non-linear r\'egime 
in which baryonic physics may have  a 
significant effect (Zhan \& Knox, 2004; White, 2004) on the power spectrum. We calculate  
non-linear effects accurately using the fitting formula of Smith et
al. (2003) for the matter power
spectrum. The use of a Gaussian likelihood is an approximation. 
The scales probed by the technique currently extend significantly into the non-linear 
r\'egime, but the integral nature of lensing is expected to make the shear more 
Gaussian (Hu \& White, 2001). However, it remains to be tested with simulations
whether the Gaussian approximation leads to significant error. 
This will be the subject of future work. Note 
that since $\bell=(\ell_x, \ell_y)$ the $\ell$ 
range used corresponds to $26$ independent $\bell$-modes.  

The use of spherical Bessel functions in the coefficients used means
that, for any given $\bell$-mode, there is a range of $k$ for which the
signal is zero up until a particular value of $k\approx |\bell|/r_{\rm
  max}$ (see Castro et al., 2005). These zero modes result in
singular covariance matrices, however this behaviour can be taken into
account using the prescription given in Appendix A.

\subsection{3D Cosmic Shear Results}
\label{3D Spectral Test Results}
This Section presents the result of applying the 3D cosmic shear
analysis to the 
CDFS and S11 fields. The results will be compared with the Fisher
matrix analysis of Heavens et al. (2006) and with the 2D cosmic shear
analysis of Brown et al. (2003). Unless otherwise stated the fiducial
cosmology that will be assumed throughout this Section is
$\Omega_m=0.3, \Omega_{de}=0.7, \Omega_b=0.04, h=0.71, \sigma_8=0.8,
w=-1.0, n_s=1.0, \alpha_n=0.0$, any constraints for
particular parameters are conditional on these values.  

\begin{figure}
\psfig{figure=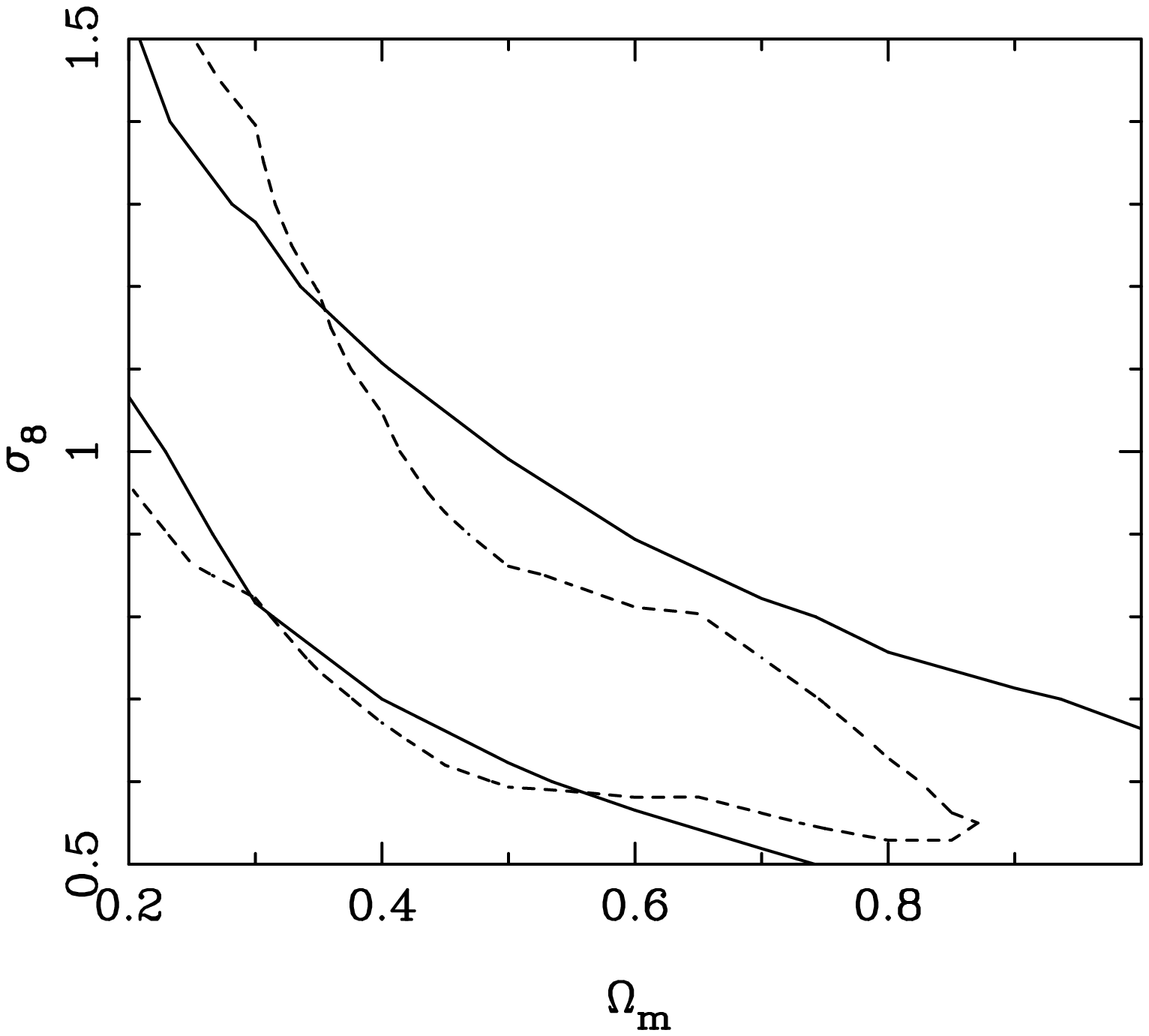,width=\columnwidth,angle=0}
\caption{The solid lines show the two-parameter $1$-$\sigma$
  conditional constraints in the ($\sigma_8$, $\Omega_m$) plane 
  from applying the 3D cosmic shear analysis to the CDFS and S11 fields
  only. The 
  dashed contours show the two-parameter $1$-$\sigma$
  conditional constraints from the Brown et al. (2003) analysis using
  $50\%$ more fields: CDFS, S11 and A901/2.}\label{fig-sig8omm}
\end{figure}

%s8, Om
Figure \ref{fig-sig8omm} shows the two-parameter $1$-$\sigma$ contours
from applying the 3D cosmic shear analysis to the CDFS and S11 fields only. The
dashed line in Figure \ref{fig-sig8omm} shows the two-parameter
$1$-$\sigma$ contours from Brown et al. (2003) where a traditional 2D
cosmic shear analysis was performed on all three COMBO-17 fields, CDFS,
S11 and A901/2 using only galaxies with accurate redshifts. It can be
seen that the 3D cosmic shear analysis constrains a very similar area in the
($\sigma_8$, $\Omega_m$) plane, particularly at the concordance values
of $\sigma_8$ and $\Omega_m$ using less than two thirds the number of
galaxies used in the 2D analysis (since the A901/2 field contains more
galaxies than the CDFS and S11 only $63\%$ of the galaxies used in the
2D analysis have been analysed). 

%\begin{figure}
%\psfig{figure=alphabeta.eps,width=\columnwidth,angle=0}
%\caption{Constraining the parameters $\alpha$ and $\beta$ in the
%  functional fit $\sigma_8(\Omega_m/0.3)^{\beta}=\alpha$. The solid
%  lines show the two-parameter $1$-$\sigma$ 
%  constraints in the ($\alpha$, $\beta$) plane 
%  from applying the 3D cosmic shear analysis to the CDFS and S11 fields
%  only. The 
%  dashed contours show the two-parameter $1$-$\sigma$
%  constraints from the Brown et al. (2003) analysis using
%  the CDFS, S11 and A901/2 fields.}\label{fig-alphabeta}
%\end{figure}

A common way to express the constraint in the ($\sigma_8$, $\Omega_m$)
plane is to constrain the parameterisation
$\sigma_8(\Omega_m/0.3)^{\beta}=\alpha$, where $\beta$ expresses the
curvature of the constraint and $\alpha$ the normalisation of the
curve. The 3D cosmic shear analysis
constrains these parameters to be 
\ba
\alpha=1.06^{+0.17}_{-0.16}\nn
\beta=0.57^{+0.19}_{-0.19} 
\ea
these are consistent with the constraints from the 2D analysis of
Brown et al. (2003). It should be noted however that 
Brown et al. (2003) have a maximum $\ell$ limit of $\sim 10,000$, so that 
they use substantially more modes in the angular direction. 
Also, their main result of
$\sigma_8(\Omega_m/0.3)^{0.49}=0.72^{+0.08}_{-0.09}$ included
galaxies with unknown redshifts, the result shown (dashed line in Figure 
\ref{fig-sig8omm}) is their result
when considering galaxies with only reliable redshift estimates. 
Since the range in $\ell$-modes and the number of galaxies are 
different in the two analyses this comparison cannot make any conclusions 
on the relative merit of the two techniques. It is sufficient to say that 
our results agree with Brown et al. (2003) when
we include the same sort of data i.e. from an analysis in which the 
majority of the galaxies were the same the results are consistent with 
one another. 

The high value of $\sigma_8$ is unexpected for
the CDFS and S11 fields, the CDFS result on it own favours a lower $\sigma_8$ (with 
substantially increased error contours), so that the cluster in the S11 field appears to
increase the most likely value of $\sigma_8$. Also the effect of not including galaxies 
with photometric redshifts appears to increase the most likely clustering value in 
Brown et al. (2003) (compare Figures 19 and 20 in Brown et al., 2003). 
Since we do not include galaxies for which a redshift is unknown 
we may expect to find a high clustering value in a similar way to 
Brown et al. (2003). We only used galaxies with reliable redshifts 
since
this paper is a proof of concept for the 3D weak lensing methods, however 
galaxies with unknown redshifts could also be included in this
analysis. 

The Fisher matrix calculations in Heavens et al. (2006) can be used to
predict the estimated uncertainties from this analysis. Fisher
matrix predictions, by construction, predict Gaussian likelihood
surfaces, the curved constraint shown here highlights one limitation
of the Fisher matrix technique to predict uncertainties when the
errors are so large. However, using the techniques outlined in Heavens
et al. (2006), we predict, for
two COMBO-17 fields, a conditional constraint of $\Delta\sigma_8=0.19$
(assuming $\Omega_m=0.3$). This is in agreement with the measured
conditional error of $\sigma_8(\Omega_m=0.3)=1.05\pm 0.20$,
highlighting that the predictions 
made in Heavens et al. (2006) are reliable. The values used in the
Fisher matrix 
calculation were an area of $A=0.52$ square degrees, to a median redshift
of $z_{\rm median}=0.8$ using the photometric redshift error given in
equation (\ref{photozeq}).   

\begin{figure}
\psfig{figure=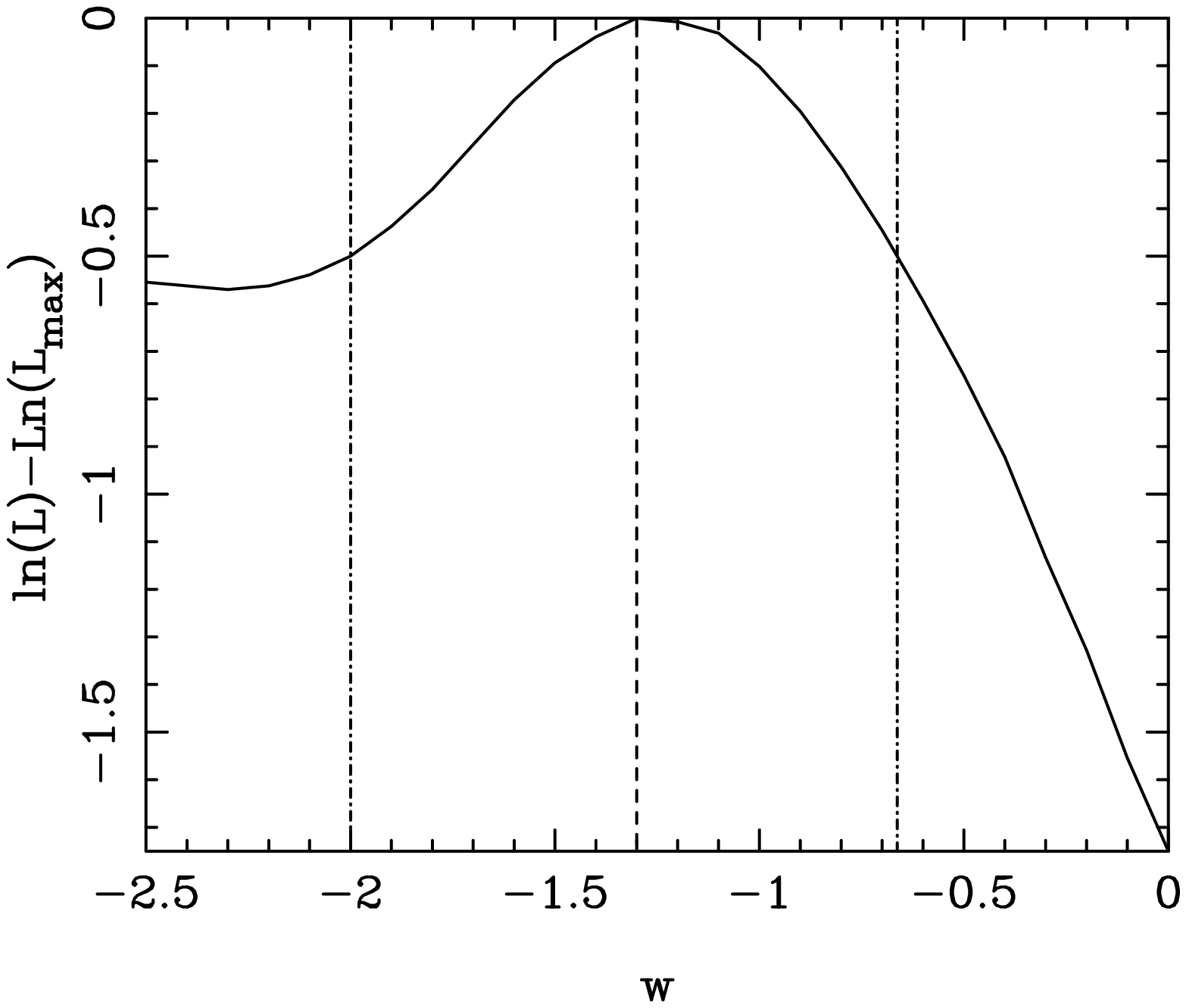,width=\columnwidth,angle=0}
\caption{The one-parameter maximum likelihood constraint on $w$ from
  the CDFS and S11 fields using the 3D cosmic shear analysis. The dashed line
  shows the most likely value and the dot-dashed show the
  one-parameter $1$-$\sigma$ constraints.}\label{fig-s11cdfw0} 
\end{figure}

Figure \ref{fig-s11cdfw0} shows the conditional constraint on $w$
from the CDFS and S11 field only using the 3D cosmic shear analysis. 
The constraint is asymmetric in that the range $w< -1$ is more
likely than $w> -1$. This is due to the fact that values of $w<
-1$ represent dark energy scenarios in which the dark energy density
is less in the past, so it is more difficult to constrain its
equation of state. Semboloni et al. (2006) also found a similar asymmetric
constraint when using weak lensing tomography applied to the CFHTLS survey. 
The conditional constraint on $w$ is 
\be
w=-1.27^{+0.64}_{-0.70}.
\ee
This result is consistent with other observations (for example
Spergel et al., 2006) and with a cosmological constant model for dark
energy. The Fisher matrix calculations, presented in Heavens et
al. (2006) predict a conditional error on $w$ from two COMBO-17
fields to be $\Delta w=0.62$ which is in agreement with the
constraints presented here. 

Typical reduced $\chi^2$ values for a given $\bell$-mode in the CDFS and S11
fields analyses are $\chi^2_{CDFS}\approx 1.01$ and $\chi^2_{S11}\approx
0.98$, the number of degrees of freedom for a given $\bell$-mode are
the corresponding number of non-singular $k$-modes used in the
analysis, typically $\sim 600$ for an average $\bell$-mode. The range
of $\chi^2$ values are consistent with a good fit to the data. 

\section{The Geometric Shear-Ratio Analysis}
\label{The Geometric Ratio Test}
We have applied the geometric shear-ratio analysis to the A901/2 field
of the COMBO-17 
survey in order to conditionally constrain $w$, and compare the
measured constraint with the predicted constraint from a Fisher
matrix calculation. The results presented
in this Section are an extension and use of the methods outlined in
Taylor et al. (2006).

\subsection{Geometric Shear-Ratio Likelihood}
To implement the geometric shear-ratio analysis, we first selected the
peaks in the convergence field of the three clusters, A901a, A901b
and A902. Taylor et al. (2004) have shown that there is a fourth
cluster, CB1, in this field, which lies behind A902 at a redshift
of $z=0.42$. Here we shall ignore the contribution of this
cluster, although this will in principle bias our results
slightly. To estimate the effect of the bias the CB1 cluster increases
the tangential shear, at $z\gs 0.4$, by $\delta\gamma_t\ls 0.02$ (see
Taylor et al., 2004). Using the simple error formula from Taylor et
al. (2006) this increase in tangential shear may bias the value of $w$
by $\delta w\ls +0.03$.

We use the positions of the centre for each cluster given by Taylor et
al. (2004), in which the tangential shear around each cluster is used
to determine the $3$D position and mass of the clusters. We averaged
the tangential  
shear in annuli around each cluster in a series of redshift bins,
following Taylor et al. (2004; see Figure 3), the width of the
redshift bins is equal to the photometric redshift error at the
redshift of the bin using the result from Section \ref{Photometric
  Redshifts}. The lensing signal from a cluster is given by the
tangential shear
\be 
\gamma_t=-[\gamma_1\cos(2\varphi)+\gamma_2\sin(2\varphi)],
\ee
for a given redshift bin. The error on the tangential shear was
estimated by the orthogonal  shear signal
\be
\gamma_{\times} = [-\gamma_1 \sin (2\varphi)+\gamma_2 \cos (2\varphi)].
\ee
Since a lensing cluster induces a tangentially aligned shear
signal, any orthogonal component is assumed to be due to noise. In this
case we define a polar coordinate system about the cluster centre and
assume a circularly-symmetric mass profile, which should produce a
purely tangential signal. Under these assumptions any orthogonal component
is taken to be due to measurement error, as opposed to deviations from
circular symmetry in the lens (which can also produce an orthogonal component in the
circular coordinate system chosen about the cluster center).
The tangential shear in each angular and redshift bin was then
fitted with a least-square fit to a singular isothermal sphere (SIS)
profile;
\be
\gamma_{t,{\rm SIS}}(\theta,z)=\frac{1}{2\theta}\theta_E(z).
\ee
$\theta_E(z)$ is the Einstein ring radius which
parameterises the amplitude of the tangential shear as a function
of source redshift.
Note that the
assumption of a SIS is not necessary, in the case of
a large data set the average tangential shear in an aperture could be
measured directly, and the orthogonal shear component for the error, with no
assumption on the radial tangential shear profile made. Since
this data 
set consists of only three small clusters the SIS was adopted so
that a signal could be measured, and for the radii from the centre of
the clusters probed should be an adequate approximation.  
 
Here $D$ denotes data and $R$ is the
theoretical estimate for the shear ratio, dependent on cosmology. The
theoretical ratio of shears, $R_{ij}$ for a pair of  redshift bins
could then be 
estimated by 
\be
R_{ij}= \frac{\theta_E(z_i)}{\theta_E(z_j)}=\frac{(r_{\rm
    lens}-r_i)/r_i}{(r_{\rm lens}-r_j)/r_j}
\ee
where $r$ is the predicted comoving distance for a given
cosmology. The data is simply the ratio of tangential shears
\be
D_{ij}=\frac{\gamma_t(z_i)}{\gamma_t(z_j)}.
\ee
The measured ratio $D_{ij}$ and the calculated ratio $R_{ij}$ are then
used in the likelihood function, summing over 
all pair-pair configurations, given by equation
\ba
\lefteqn{ -2 \ln L_c(\Omega_{de},\Omega_m,w,w_a|\D) =} \nn
& & \hspace{2.cm}  \sum_{\mu,\nu} (R_\mu - D_\mu ) [C^{RR}_{\mu
    \nu}]^{-1}
(R_\nu - D_\nu),
\label{like}
\ea
for a given cluster. The notation for a pair of background
bins has been compressed as
$\mu=(i,j)$ and $\nu=(m,n)$, and all degenerate pair-pair combinations
have been accounted for. The likelihood
functions for multiple clusters are multiplied. We can write the
covariance matrix for shear ratios as 
\be
C_{\nu \mu}^{RR} \equiv \lgl \Delta R_\nu\Delta R_\mu \rgl.
\ee
The full covariance matrix includes shot noise and cosmic shear terms.
For a full description see Taylor et al. (2006).

\begin{figure}
 \psfig{figure=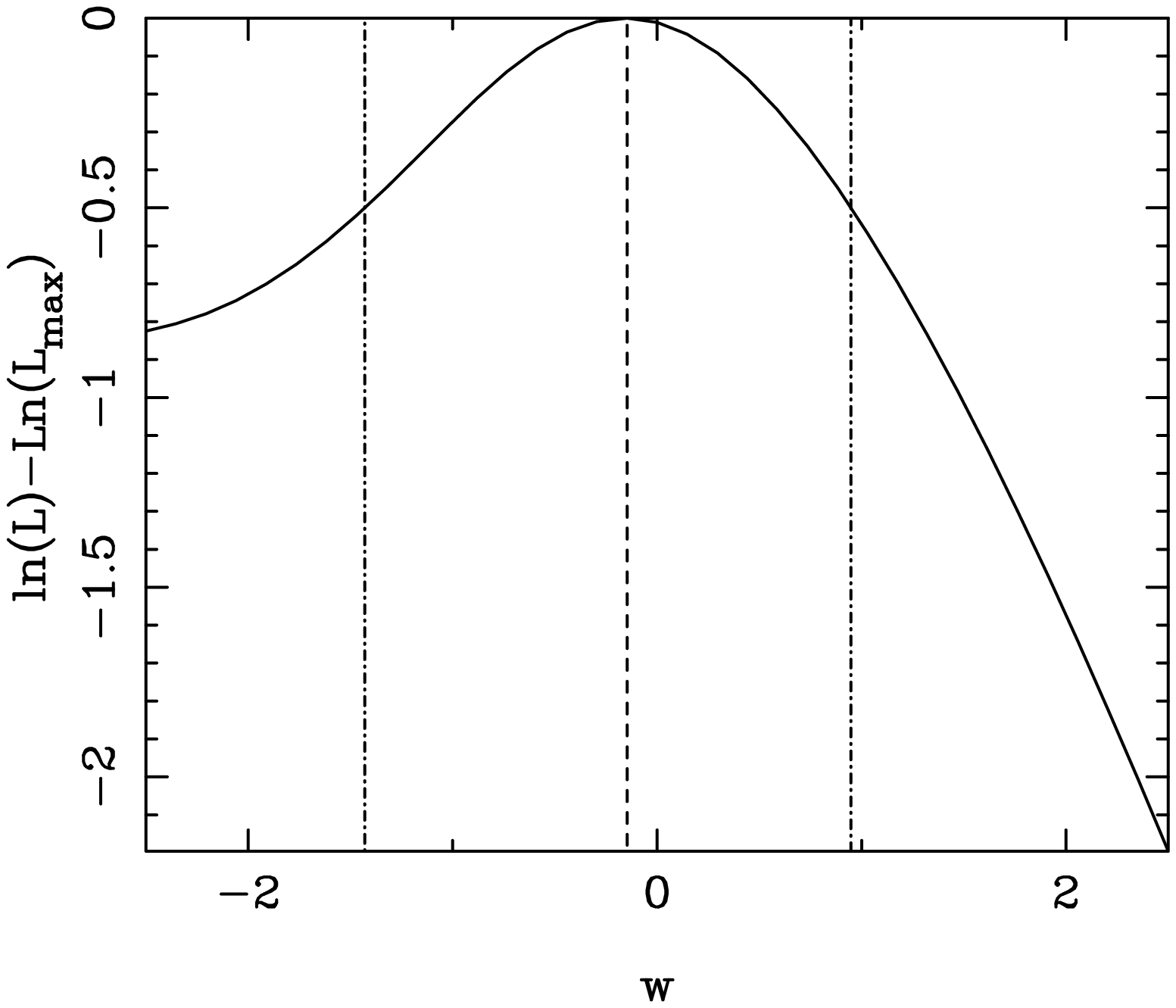,width=\columnwidth,angle=0}
 \caption{The dark energy geometric shear-ratio analysis applied to the supercluster
 Abell A901/2. The dashed line marks the maximum likelihood value, the
 dot-dashed lines show the one-parameter $1$-$\sigma$ limits. Note
 that the x-axis scale has been extended relative to Figures
 \ref{fig-s11cdfw0} 
 and \ref{combo-dark_energy} to encompass the confidence
 limits of this analysis.}
 \label{combo-dark_energy}
\end{figure}

\subsection{Geometric Shear-Ratio Results}
The results are shown as a 1D likelihood plot in Figure
\ref{combo-dark_energy}. The result is conditional on $\Omega_m=0.30$
, $\Omega_{de}=0.70$ and $w_a=0.0$. 
The dashed line is our measured most likely value, the dot-dashed
lines are the one-parameter 
$1$-$\sigma$ (68\%) confidence limits. Our constraint on $w$ is:
\be
w= -0.11^{+1.05}_{-1.29}.
\ee
The constraint again shows an asymmetry between the $w<-1$ and $w>-1$
regions for the same reason given in the 3D cosmic shears constraint,
that $w<-1$ represents a lower dark energy density in the past.
The minimum $\chi^2$ value is $\chi^2_{\rm min}=122$ which is consistent
with the number of degrees of freedom in the experiment. Given that
$z_{\rm max}\approx 2.0$ and $\Delta z\approx 0.05$ and we analyse
$N_{\rm cluster}=3$ clusters the predicted $\chi^2_{\rm min}=(z_{\rm max}/\Delta
z)N_{\rm cluster}\approx 120$ so that $\chi^2_{\rm reduced}=1.01$
and should be $\chi^2_{\rm reduced}=1\pm 0.12$.

The result is consistent with other constraints on $w$, and the
confidence limits allow for most dark energy models. It should be
emphasised that this constraint comes from only three small
clusters. 

The Fisher matrix calculations, in Taylor et al. (2006)
predict a conditional constraint on 
$w$ of $\Delta w=1.10$ for COMBO-17 which was created by assuming
only three
clusters at $z=0.16$ with ${\rm M}_{\rm cluster}=10^{14}{\rm M}_{\odot}$. The
predicted conditional constraint is approximately the same as the
measured constraint, thus verifying the Fisher matrix methodology. 

\section{A Combined Constraint on \lowercase{$w$}}
\label{A Combined Constraint}
\begin{figure}
\psfig{figure=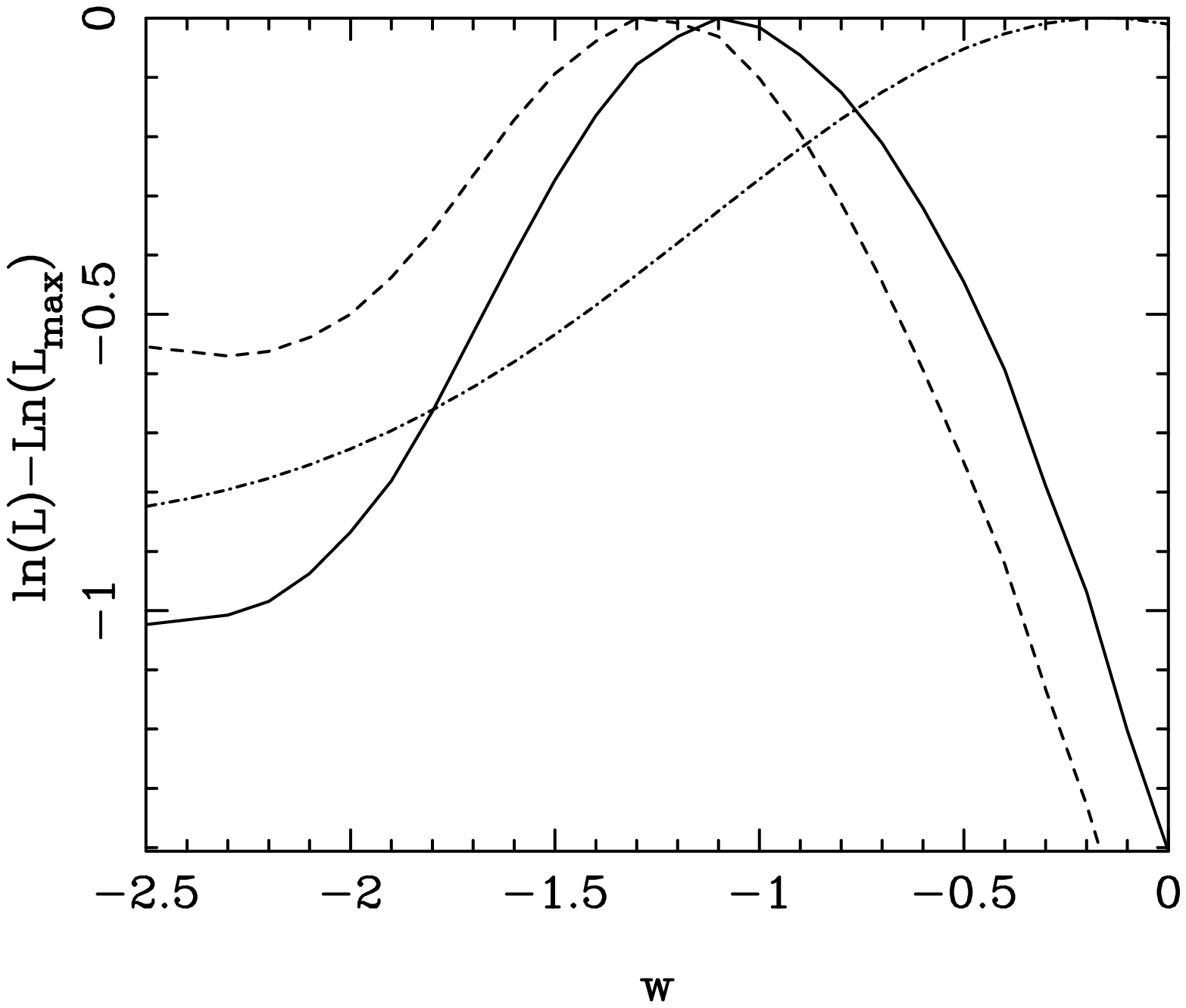,width=\columnwidth,angle=0}
\caption{The one-parameter maximum likelihood constraint on $w$
  obtained by combining the geometric shear-ratio analysis constraint
  from A901/2 field 
  and the 3D cosmic shear analysis constraint from the CDFS and S11 fields. The
  solid line  
  shows the combined constraint, the dashed line shows the 3D cosmic
  shears constraint shown in Figure \ref{fig-s11cdfw0}, the dot-dashed
  line shows the geometric shear-ratios constraint shown in Figure
  \ref{combo-dark_energy}.}\label{fig-combine}
\end{figure}

%combine geo and spec w0's
Since the A901/2 field was analysed separately from the CDFS and S11
fields the geometric shear-ratio analysis constraint can be combined with the constraint
from the 3D cosmic shear analysis. Figure \ref{fig-combine} shows the result of
adding the constraints shown in Figures \ref{fig-s11cdfw0} and
\ref{combo-dark_energy} from the CDFS 
and S11 fields and the A901/2 field respectively. 
The resulting conditional constraint on $w$ is 
\be 
w=-1.08^{+0.63}_{-0.58}.
\ee
This result demonstrates the value of combining the two techniques
that can analyse distinct parts of the data. A region of particularly
high density such as the A901/2 field would raise issues of sample
bias in a cosmic shear/spectral approach, however the geometric 
shear-ratio analysis necessarily needs such areas. The effect of
adding the geometric shear-ratio analyses constraint is 
the most likely 
value becoming more positive, and a slight reduction in the error. The
most likely value of $w=-1.08$ is in complete agreement with other
observations (for example Semboloni et al, 2006; Spergel et al.,
2006) and is close to the value of $w$ expected if dark energy is a
cosmological constant. The caveat on this conclusion is that it is
conditional on the other cosmological parameters being fixed, and the
error on $w$ is still fairly large. Finally it is surprising, given
the size of the error, that the maximum likelihood estimate is so
close to $w=-1$. 

\section{Conclusion}
\label{Conclusion}
%What have we done?
In this paper we have applied the 3D weak lensing methods introduced
in Heavens (2003) and Jain \& Taylor (2003) and developed in
Heavens et al. (2006) and Taylor et al. (2006) to data for the first
time. We used the COMBO-17 data set, a multi-band photometric survey,
ideal for a weak lensing study. We used three of the fields, CDFS, S11
and A901/2 
to place conditional constraints on the equation of state of dark
energy, $w$ and the ($\sigma_8$, $\Omega_m$) plane. The size of the
dataset is small so we did not expect accurate constraints and this
paper is essentially a proof of concept for the methods. To this end
we compute conditional constraints and compare with Fisher matrix
predictions. A full analysis of a larger data set should, of course,
quote marginal errors. 

The first method used
was the 3D cosmic shear analysis which uses
spherical harmonics to describe the fully 3D shear field. Applying the
3D cosmic shear analysis to the CDFS and S11 field we conditionally
constrained an 
area in the ($\sigma_8$, $\Omega_m$) plane described by
$\sigma_8(\Omega_m/0.3)^{0.57\pm 0.19}=1.06^{+0.17}_{-0.16}$. This is in
agreement with Brown et al. (2003) in which a 2D cosmic shear 
analysis was performed on $50\%$ more data. The application of 3D
cosmic shear
conditionally constrained the equation of state of dark energy to
$w=-1.27^{+0.64}_{-0.70}$. 

The second method used was the geometric shear-ratio analysis, which
takes the ratio of the tangential shear around galaxy 
clusters at different redshifts. We applied this analysis to the A901/2
field which contains three small clusters conditionally constraining
$w=-0.15^{+ 1.07}_{-1.28}$. 

Combining the constraint on
$w$ from the geometric shear-ratio analyses application to the A901/2
field and the 
3D cosmic shear constraint from CDFS and S11 we conditionally constrain
$w=-1.08^{+0.63}_{-0.58}$. For discussions on the relative
merit of the two methods and varying observing strategies see Heavens
et al. (2006) and Taylor et al. (2006).  These papers also discuss the
effects of systematics, number of observing bands
used and the effects of the assumed fiducial cosmology. 

%Interesting Constraints??
The constraints presented here do not improve much on
our cosmological understanding, they are however in agreement with the
currently accepted concordance model of Spergel et al. (2006). The
constraint on $w$ from such a small data set is encouraging and, with
the warning and caveat that it is a conditional error, it is
\emph{a forteriori} consistent with dark energy being a cosmological constant.

%Actual Marginal Constraints.--> why these values??
In order for these results to become more complete we could marginalize
over an increasingly large cosmological parameter set. However this
would rapidly result in a loss of any constraint, for such a small
survey area, due to degeneracies
between the parameters as shown in Heavens et al. (2006) and Taylor et
al.(2006). The result errors agree with the Fisher matrix predictions
and are a very reliable proof of concept for these methods. 

%Fisher Matrix comparisons.
The agreement of the results presented here with the Fisher matrix
predictions using the methods presented in Heavens et al. (2006) and
Taylor et al. (2006) are a validation of the Fisher matrix framework
and an encouraging sign that the predictions made in these papers are
robust and accurate. COMBO-17 was an ideal survey upon which to test
these 3D weak lensing methods, however with much larger area surveys
with fewer observing bands, such as VST-KIDS, Pan-STARRS (Kaiser et al.,
2005) or the Dark Energy Survey (Wester, 2005), these techniques could
constrain the dark energy equation of state to approximately $1\%$ at
a particular redshift.  

\section*{Acknowledgments}

We would like to thank Patricia Castro for helpful
discussions. TDK acknowledges a PPARC studentship. 
MLB acknowledges the support of a PPARC Fellowship.
CW was supported by a PPARC Advanced Fellowship.

\onecolumn

\section*{Appendix A: Removal of Singular Modes}

In this Appendix we show how any singular ($k$, $\ell$) modes can be
removed from the covariance matrices used in the 3D cosmic shear
likelihood analysis.

We begin with a square covariance matrix $C$ which can be
decomposed using a standard singular value decomposition (SVD) into
\be 
\label{AAeq01}
C=UWV^T,
\ee
where $W$ is a diagonal matrix that contains the singular values. Note
that $U$ and $V$ are eigenvector matrices of $CC^T$ and that
$U^{-1}=U^T$ and $V^{-1}=V^T$. Our covariance matrices are symmetric
so that $U=V$ in this case. Now consider one of our data vectors
$\hat\gamma^A_i(k,\bell)$ represented by $\ux$ which can be 
transformed to a new data vector $\uy$ via
\be
\label{AAeq02}
\uy=B\ux
\ee
where $B$ can be any, not necessarily square, transformation matrix. A new
covariance can then be defined
\be
C'_{ij}\equiv\langle y_i y_j\rangle=\langle B_{ik}x_k
B_{jl}x_l\rangle=B_{ik}B_{jl}C_{kl},
\ee
which implies that 
\be
\label{AAeq03}
C'=BCB^T.
\ee

The choice of transformation, in this case, is motivated by decomposing
$C$ using a Cholesky 
decomposition which yields $C=UWU^T=LL^T$, where $L=UW^{1/2}$. So we
use $B=\tilde{W}^{-1/2}U^{-1}$ where $\tilde{W}^{-1/2}=W^{-1/2}$
except that the elements of the inverse $W$ matrix $1/w_i$ have been
replaced with zero if 
$(1/w_i)\geq$(threshold) where the threshold represents machine
precision (see Press, 1993; Numerical Recipes). The matrix $B$
now contains a band 
of values below which 
zeros remove any singular modes from either the data vector or the
covariance matrix via equations (\ref{AAeq02}) and (\ref{AAeq03}). 

The transformation is performed using a fiducial cosmology
(the choice of this does not affect the results) to yield a
transformation matrix $B$ which is then used throughout. $\uy$ and $C'$
replace $\ux$ and $C$ in the likelihood analysis. 

\end{document}